\documentclass[12pt, a4paper]{article}
\usepackage[utf8]{inputenc}
\usepackage[T2A]{fontenc}
\usepackage[english]{babel}

\usepackage{feynmf}
\usepackage{tabularx}
\usepackage{booktabs}
\usepackage{footnote}

\usepackage{indentfirst}
\usepackage{amsmath, amssymb, mathrsfs}

\usepackage{graphicx}

\usepackage{float}

\usepackage{caption}
\DeclareCaptionLabelSeparator{Sep}{\bf{:} }
\usepackage{csquotes}

\usepackage{array}
\usepackage{multirow}
\usepackage{hhline}

\usepackage{csquotes}
\usepackage{textcase}
\usepackage{xcolor}

\usepackage{hyperref}
\hypersetup{colorlinks,urlcolor=blue,linkcolor=blue,citecolor=blue}

\usepackage{geometry}
\usepackage{setspace}
\newcommand{\zzllvv}{\ensuremath{{ZZ\rightarrow\ell\ell\nu\nu}}}
\newcommand{\wzlvll}{\ensuremath{{WZ\rightarrow\ell\nu\ell\ell}}}
\newcommand{\wwlvlv}{\ensuremath{{WW\rightarrow\ell\nu\ell\nu}}}
\newcommand{\zllj}{\ensuremath{{Z(\rightarrow\ell\ell)\text{ + jets}}}}
\newcommand{\ztata}{\ensuremath{{Z\rightarrow\tau\tau}}}
\newcommand{\zzllll}{\ensuremath{{ZZ\rightarrow 4\ell}}}
\newcommand{\cgp}{\ensuremath{C_{G+}/\Lambda^4}}
\newcommand{\cgm}{\ensuremath{C_{G-}/\Lambda^4}}
\newcommand{\cbtw}{\ensuremath{C_{\tilde{B}W}/\Lambda^4}}
\newcommand{\cbw}{\ensuremath{C_{BW}/\Lambda^4}}
\newcommand{\cbb}{\ensuremath{C_{BB}/\Lambda^4}}
\newcommand{\cww}{\ensuremath{C_{WW}/\Lambda^4}}
\newcommand{\ogp}{\ensuremath{\mathcal{O}_{G+}}}
\newcommand{\ogm}{\ensuremath{\mathcal{O}_{G-}}}
\newcommand{\obtw}{\ensuremath{\mathcal{O}_{\tilde{B}W}}}
\newcommand{\obw}{\ensuremath{\mathcal{O}_{BW}}}
\newcommand{\obb}{\ensuremath{\mathcal{O}_{BB}}}
\newcommand{\oww}{\ensuremath{\mathcal{O}_{WW}}}
\newcommand{\lagr}{\ensuremath{\mathcal{L}}}
\newcommand{\oper}{\ensuremath{\mathcal{O}}}
\newcommand{\amp}{\ensuremath{\mathcal{A}}}
\newcommand{\mmu}{\ensuremath{\boldsymbol{\mu}}}
\newcommand{\tth}{\ensuremath{\boldsymbol{\theta}}}
\newcommand{\htth}{\ensuremath{\hat{\tth}}}
\newcommand{\hhtth}{\ensuremath{\hat{\hat{\hspace{-0.5mm}\tth}}}}
\newcommand{\hmmu}{\ensuremath{\hat{\mmu}}}

\makeatletter
\def\maketitle{
\begin{center}\textbf{\Large{\@title}}
\par\vspace{2mm}\textbf{\@author}
\begin{singlespace}
\par$^a$National Research Nuclear University MEPhI, Moscow, Russia
\par$^b$A. Alikhanyan National Science Laboratory (Yerevan Physics Institute), Yerevan, Armenia\par
*E-mail: AESemushin@mephi.ru\par
$^\dag$E-mail: EYSoldatov@mephi.ru
\end{singlespace}\end{center}
}
\makeatother

\title{Composite effective field theory signal in case of searching for neutral triple gauge couplings with $\boldsymbol{\zzllvv}$ production}
\author{Artur E. Semushin$^{a,b,}$*, Evgeny Yu. Soldatov$^{a,\dag}$}
\date{}

\begin{document}
\maketitle
\begin{abstract}
    In this work manifestations of physics beyond the Standard Model are parameterized with higher-dimension operators and Wilson coefficients using effective field theory. In order to set more stringent experimental limits on the Wilson coefficients it is crucial to use new methods of sensitivity increasing that work independently with luminosity growth. Method of composite anomalous signal allows one to set the limits on Wilson coefficients more precisely and stringent by accounting for impact of EFT operators on background processes in addition to the conventional anomalous contribution from the signal process. This work presents aforementioned methodology applied to the $\zzllvv$ production, that is sensitive to the neutral triple couplings of gauge bosons $Z$ and $\gamma$. It is found that the main background BSM contribution comes from $\wzlvll$ background. Improvement of the limits on the Wilson coefficients due to this background depends on the coefficient and is up to 57.5\% (59.1\%) for one- (two-) dimensional limits in linear~+~quadratic effective field theory model and up to 94.4\% for one-dimensional limits in linear effective field theory model.
\end{abstract}

\newpage
\section{Introduction}
The Standard Model (SM) of particle physics combines knowledge about all known elementary particles and their interactions. The first observation of the Higgs boson has been made in 2012 at the Large Hadron Collider (LHC)~\cite{ATLAS:2012yve, CMS:2012qbp} and become a very strong proof of the correctness of the SM. Furthermore, the SM was precisely tested in high-energy-physics experiments by studying rare phenomena. These studies yielded in a good agreement between theoretical predictions and experimental data~\cite{ParticleDataGroup:2022pth}.

However, the SM has internal inconsistencies, such as dependence on a set of free parameters and fine-tuning problem. Moreover, the SM do not describe gravity and some observed phenomena. For instance, it fails to describe neutrino oscillations and baryon asymmetry of the Universe. Due to the aforementioned facts, the SM should be considered as an effective theory, that is low-energy approximation of the more general theory. Therefore, experimental tests of the SM are usually accompanied by searches for beyond-the-SM (BSM) phenomena, so-called new physics, and its observation is one of the main goals of the modern high-energy-physics experiments.

New physics can be searched for in two ways. Direct approach implies search for new particles, whereas indirect approach is based on looking for deviations from the SM in the interactions of already known particles. Since the evidences of new particles presence at the currently available energies was not found~\cite{Rappoccio:2018qxp, Mitsou:2023dgf}, the indirect way becomes more prospective. The searches for anomalous couplings refer to this way and allow one to look for manifestations of heavy new physics at currently accessible energies. A convenient way to parameterize the anomalous couplings is an effective field theory (EFT)~\cite{Weinberg:1978kz, Degrande:2012wf}, that uses higher-dimension operators. Free parameters of this approach, Wilson coefficients, can be constrained experimentally. Non-zero Wilson coefficients affect all physical processes, and experimental data is a composite of contributions from signal and background processes. Therefore, it is important to study the impact of EFT on background processes in addition to the conventional impact of EFT on signal process, since background EFT contributions can significantly change the experimental sensitivity~\cite{Symmetry2022}.

EFT impact on background processes strongly depends on the channel being studied and the choice of EFT operator basis. In this work, EFT contributions from backgrounds are studied on the example of $\zzllvv$ production at the LHC. Studies of this process are usually interpreted in terms of anomalous neutral triple gauge couplings (nTGCs) due to the large sensitivity~\cite{ATLAS:2019xhj, CMS:2015qgb}. Therefore, dimension-eight operators affecting nTGCs are used in this work.

\section{Effective field theory framework}
EFT is a powerful tool to extend the SM in a model-independent way. It parameterizes manifestations of new physics with heavy energy scale $\Lambda$ at currently accessible energies as anomalous couplings. For this purpose, EFT uses operators of dimension greater than four, which is the dimension of the SM Lagrangian $\lagr_\text{SM}$. The resulting EFT Lagrangian can be written as
\begin{equation}
    \lagr = \lagr_\text{SM} + \sum\limits_{d>4} \sum\limits_i \frac{C_i^{(d)}}{\Lambda^{d-4}} \oper_i^{(d)},
\end{equation}
where $\oper_i^{(d)}$ is the $i$-th dimension-$d$ operator, and $C_i^{(d)}/\Lambda^{d-4}$ is the corresponding Wilson coefficient. The operators are constructed out of the SM fields and required to respect the SM gauge symmetries. Dimension-odd operators necessarily contain fermions and, therefore, are not valid for studying anomalous bosonic couplings. In the SM all the Wilson coefficients are zero, whereas their non-zero values imply the presence of new physics. This new physics may arise from new heavy particles interacting with the SM particles and, therefore, effectively changing the couplings between the SM particles with energies lower than the new physics scale. Integrating the new heavy fields out from the Lagrangian of new physics model, one can obtain a model-dependent low-energy effective Lagrangian (for example, Euler--Heisenberg Lagrangian~\cite{Heisenberg:1936nmg}). Wilson coefficients can be reinterpreted in terms of SM parameters and new particles' masses by matching model-independent and model-dependent Lagrangians~\cite{Quevillon:2018mfl, Remmen:2019cyz}. Therefore, limits on Wilson coefficients can be transformed into the limits on new particles parameters. This can constrain some new physics models, and it is crucial to set experimental limits on Wilson coefficients correctly and to make them more stringent if it is possible.

Neutral triple gauge couplings, i.e. $ZZZ$, $ZZ\gamma$, $Z\gamma\gamma$ and $\gamma\gamma\gamma$, are zero in the SM. They can be added to the theory Lagrangian using EFT. Dimension-six operators do not predict such couplings, therefore, dimension-eight operators are used to study nTGCs. For the study, carried out in this work, the following six dimension-eight operators are considered:
\begin{align}
    & \oper_{G\pm} = g^{-1} \tilde{B}_{\mu\nu} W^{a\mu\rho} \left( D_\rho D_\lambda W^{a\nu\lambda} \pm D^\nu D^\lambda W^a_{\lambda\rho} \right), \label{eq:ogpm} \\
    & \obtw = i \Phi^\dag \tilde{B}_{\mu\nu} \hat{W}^{\mu\rho} \{ D_\rho , D^\nu \} \Phi + \text{h.c.}, \label{eq:obtw} \\
    & \obw = i \Phi^\dag B_{\mu\nu} \hat{W}^{\mu\rho} \{ D_\rho , D^\nu \} \Phi + \text{h.c.}, \label{eq:obw} \\
    & \obb = i \Phi^\dag B_{\mu\nu} B^{\mu\rho} \{ D_\rho , D^\nu \} \Phi + \text{h.c.}, \label{eq:obb} \\
    & \oww = i \Phi^\dag \hat{W}_{\mu\nu} \hat{W}^{\mu\rho} \{ D_\rho , D^\nu \} \Phi + \text{h.c.} \label{eq:oww}
\end{align}
This basis consists of two pure gauge operators~\cite{Ellis:2020ljj, Ellis:2022zdw} and four mixed Higgs-gauge operators~\cite{Degrande:2013kka}. The former operators are classified under the assumption of $CP$-conservation, whereas the latter ones form a full basis only under the assumption that final-state $Z$ is on-shell. All these operators violate $C$-symmetry. $CP$-symmetry is conserved by $\ogp$, $\ogm$, $\obtw$ and violated by $\obw$, $\obb$, $\oww$ operators. Each operator affects a set of vertices, changing corresponding vertex functions compared to the SM ones. Effect of considered six operators on different triple gauge-boson couplings is summarized in Table~\ref{tab:EFTOperators}.
\begin{table}[h!]
    \centering
    \caption{Effect of dimension-eight EFT operators on triple gauge couplings; affected couplings are marked with $\circ$.}
    \label{tab:EFTOperators}
    \begin{tabular}{|c|c|c|c|c|}
        \hline
        Operator & $ZZZ$, $ZZ\gamma$, $Z\gamma\gamma$ & $\gamma\gamma\gamma$ & $WWZ$ & $WW\gamma$ \\ \hline
        $\mathcal{O}_{G\pm}$ & $\circ$ & $\circ$ & $\circ$ & $\circ$ \\ \hline
        $\mathcal{O}_{\tilde{B}W}$ & $\circ$ & & $\circ$ & $\circ$ \\ \hline
        $\mathcal{O}_{BW}$ & $\circ$ & & $\circ$ & $\circ$ \\ \hline
        $\mathcal{O}_{WW}$ & $\circ$ & & $\circ$ & \\ \hline
        $\mathcal{O}_{BB}$ & $\circ$ & & & \\ \hline
    \end{tabular}
\end{table}

Usually the Lagrangian is parameterized by one or two operators at a time, that yields in one- or two-dimensional constraints on the Wilson coefficients. In case if the Lagrangian is parameterized by one operator and if the process contains only one anomalous vertex, process squared amplitude and, therefore, cross section contains three terms:
\begin{equation}
    |\amp|^2 = |\amp_\text{SM} + (C/\Lambda^4) \amp_\text{BSM}|^2 = |\mathcal{A}_\text{SM}|^2 + (C/\Lambda^4) 2 \text{Re} \amp_\text{SM}^\dag \amp_\text{BSM} + (C/\Lambda^4)^2 |\amp_\text{BSM}|^2. \label{eq:EFTdecomp}
\end{equation}
The first term comes purely from the SM and does not depend on the Wilson coefficient. The second term is interference between SM and new physics contribution, so-called interference (linear) term. The third term is the quadratic term, that comes purely from new physics. In the case of two-dimensional parameterization, the process squared amplitude contains six terms. In addition to the terms that are analogous to the ones from one-dimensional parameterization, there is a cross-term, representing interference between two operators and proportional to the product of two Wilson coefficients. This term can significantly change the sensitivity.

Since energy scale $\Lambda$ is heavy, the quadratic and cross terms ($\propto\Lambda^{-8}$) should be suppressed compared to the interference terms ($\propto\Lambda^{-4}$). Moreover, interference between dimension-twelve operators and SM is also $\propto\Lambda^{-8}$, that can lead to the change of the sensitivity due to the statistical correlation between dimension-eight and dimension-twelve Wilson coefficient. However, for $CP$-odd operators interference is zero without accounting for special $CP$-sensitive variables, and for $CP$-even operators effect of interference suppression due to the allowed boson polarizations is presented~\cite{Degrande:2013kka}. Therefore, in this work a model with all terms from Eq.~\eqref{eq:EFTdecomp} (linear~+~quadratic model) is baseline and used under the assumption that interference contributions from dimension-twelve operators are suppressed compared to the quadratic contributions from dimension-eight operators.

Additionally, it is needed to show effect of background processes on linear term, since at the current experimental sensitivity quadratic term dominates. For this purpose, a model with dropped quadratic term (linear model) is used for 1D limits on Wilson coefficients on $CP$-even operators only.

\section{Modelling of the physical processes}
\begin{figure}[h!]
    \centering
    \includegraphics[width=0.7\textwidth]{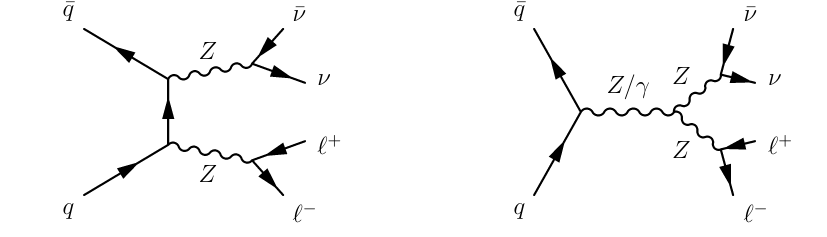}
    \caption{Examples of the Feynman diagrams for $\zzllvv$ production via the SM vertices~(left) and via a neutral triple gauge coupling~(right).}
    \label{fig:diagrams}
\end{figure}

EFT impact on background processes is studied for $\zzllvv$ production in $pp$ collisions at $\sqrt{s}=13$~TeV. Integrated luminosity of 140 fb$^{-1}$, corresponding to the data collected by the ATLAS experiment during LHC Run II~\cite{ATLAS:2022hro}, is used for the calculations. Examples of the Feynman diagrams of this process production in the SM and via nTGC are shown in Fig.~\ref{fig:diagrams}. The main background for this process is $\wzlvll$ production, where one lepton is either not identified or a $\tau$, decaying to jets or non-identified charged lepton and neutrinos. Sizeable background comes from non-resonant $\ell\ell$ production, and in this work non-resonant background is a combination of $\wwlvlv$, $\ztata$, $tt$ and $tW$ processes production. Possible mismeasurement of jet energies leads to the fake missing transverse energy, that emulates neutrinos. This background dominantly comes from $\zllj$ production. Additional background is $\zzllll$ production, where only two charged leptons are identified by the detector. Other backgrounds are not accounted, since they yield small contribution~\cite{ATLAS:2019xhj}, whereas this work uses only toy model in order to study EFT impact on background processes.

All aforementioned backgrounds are affected by non-zero Wilson coefficients of operators from Eqs.~\eqref{eq:ogpm}--\eqref{eq:oww} at the tree level. However, some backgrounds require one ($tW$, $\zllj$) or two ($\ztata$, $tt$) additional jets for being produced via triple gauge coupling. BSM contributions from such backgrounds should be small, but they are accounted in this study. The exception is impact of $\obb$ operator on processes $\wzlvll$, $\wwlvlv$ and $tW$, that is zero at the leading order, as it can be seen in Table~\ref{tab:EFTOperators}. $\obb$ affects these processes under the requirement of two additional jets, however in this case anomalous quartic gauge couplings and Feynman diagrams with two anomalous vertices are presented. Thus, EFT contributions from $\wzlvll$, $\wwlvlv$ and $tW$ productions for $\obb$ operator are assumed to be negligible.

All the processes were modelled using Monte Carlo event generator \texttt{MadGraph5\_aMC@NLO}~\cite{Alwall:2014hca, Frederix:2018nkq}. In order to model the BSM contributions, a decomposition technique in \texttt{MadGraph5\_aMC@NLO} was used, that allows generating the events, corresponding to the different squared amplitude terms, separately~\cite{PHAN2021}. Parton distribution functions, calculated by the NNPDF collaboration~\cite{NNPDF:2014otw}, were used in the computations. Parton shower, hadronization and underlying event was simulated using \texttt{Pythia8}~\cite{Sjostrand:2014zea}. For some SM processes accounting the events with additional jets significantly changes the cross section. So, \texttt{Pythia8} was also used to merge matrix element and parton shower for such processes with CKKW-L scheme~\cite{Lonnblad:2001iq}. Finally, interaction of particles with a typical LHC detector were simulated using \texttt{Delphes3}~\cite{deFavereau:2013fsa}. Note that in Monte Carlo simulation $\ell$ for $\zllj$ background means $e$ or $\mu$, whereas for other processes $\ell$ includes also $\tau$.

Object reconstruction and event selection criteria in this work are based on the study of $\zzllvv$ production with 36.1 fb$^{-1}$ of $pp$-collisions at $\sqrt{s}=13$~TeV by the ATLAS collaboration~\cite{ATLAS:2019xhj}. Some criteria are changed in order to simplify the toy model and since this study does not use full simulation of the detector. The criteria of the event selection are summarized in Table~\ref{tab:selection} \footnote{This work uses right-handed coordinate system with origin at the centre of the detector, $x$- and $y$-axes directed to the centre of the LHC and upwards, and $z$-axis located along the beam pipe. Transverse momentum $\vec{p}_\text{T}$ is calculated in the $xy$ plane. Variable $\Delta R = \sqrt{(\Delta \eta)^2 + (\Delta \varphi)^2}$ measures angular distance between two objects, where $\eta = - \log \tan (\theta/2)$ is the pseudorapidity.}. After applying these criteria, normalization of the SM $\zllj$ background was found too large. It can be explained by the fact that this background is modelled badly and usually estimated using data-driven techniques~\cite{Kazakova:2023cbr}, and detector simulation in \texttt{Delphes3} reproduces this background incorrectly. So, for the SM $\zllj$ background the shape was taken from Monte Carlo modelling, and the normalization was taken from the ATLAS study~\cite{ATLAS:2019xhj}. Moreover, for non-resonant backgrounds both shape and normalization were modelled badly due to the low statistics of Monte Carlo samples. Therefore, for this background shape was taken the same as for $\zzllvv$ production, and normalization was also taken from the ATLAS study~\cite{ATLAS:2019xhj}.
\begin{table}[h!]
    \centering
    \caption{Event selection criteria.}
    \label{tab:selection}
    \begin{tabular}{|c|}
    \hline
    Two leptons of same flavor and opposite signs \\
    $p_\text{T}^{\ell_1}>30$ GeV, $p_\text{T}^{\ell_2}>20$ GeV \\
    $p_\text{T}^{\ell\ell}>150$ GeV \\
    $76<m_{\ell\ell}<106$ \\
    $\Delta R _{\ell\ell} > 1.9$ \\
    $E_\text{T}^\text{miss}>120$ GeV \\
    $N_j\geq 0$, $N_{b\text{-jet}}=0$ \\
    $\left|\sum\limits_\text{objects} \mathbf{p}_\text{T}^{~j}\right| / \sum\limits_\text{objects} p_\text{T}^j > 0.65$ \\ \hline
    \end{tabular}
\end{table}

\begin{figure}[h!]
\centering
\includegraphics[width=0.3\textwidth]{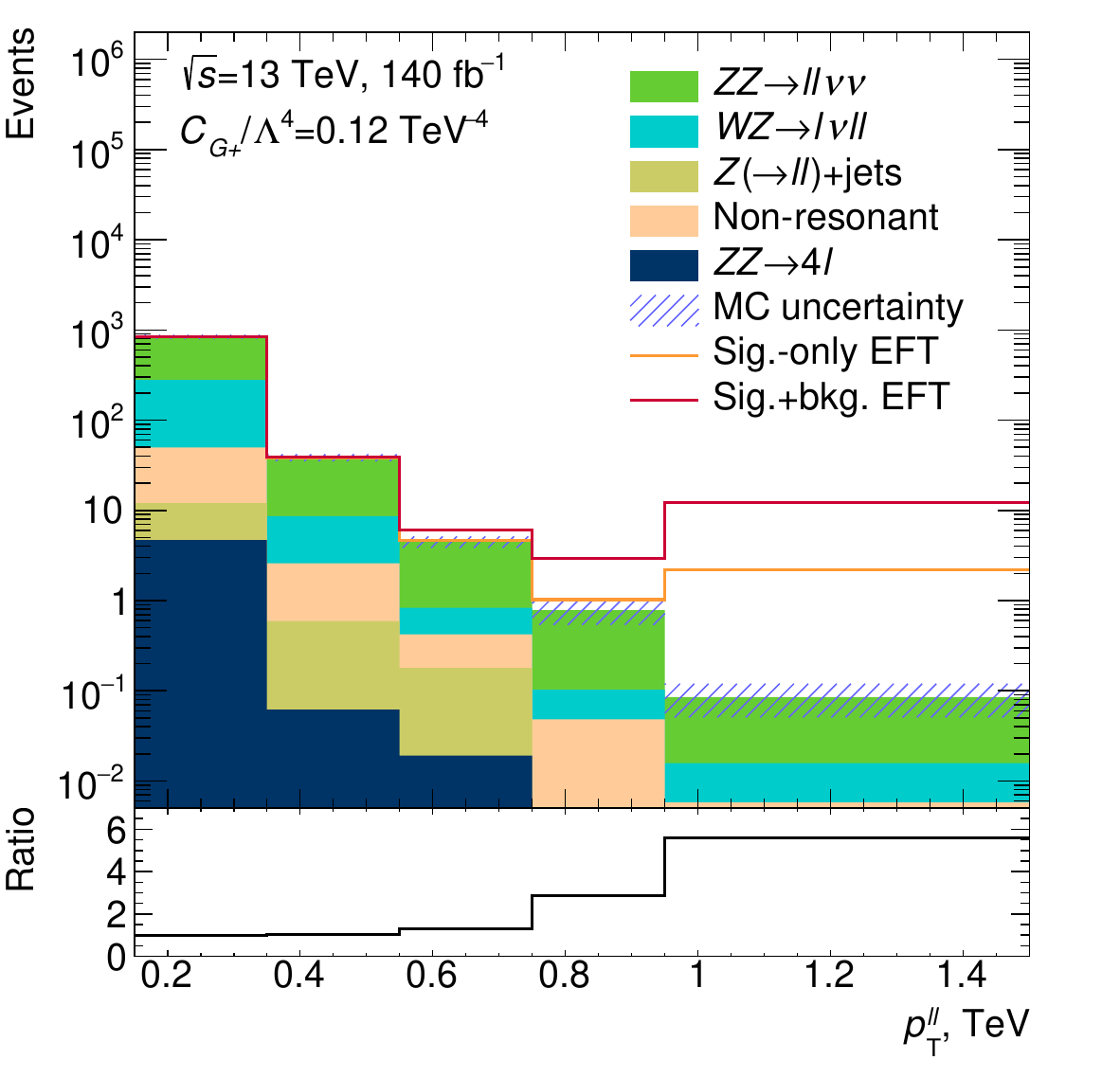}
\includegraphics[width=0.3\textwidth]{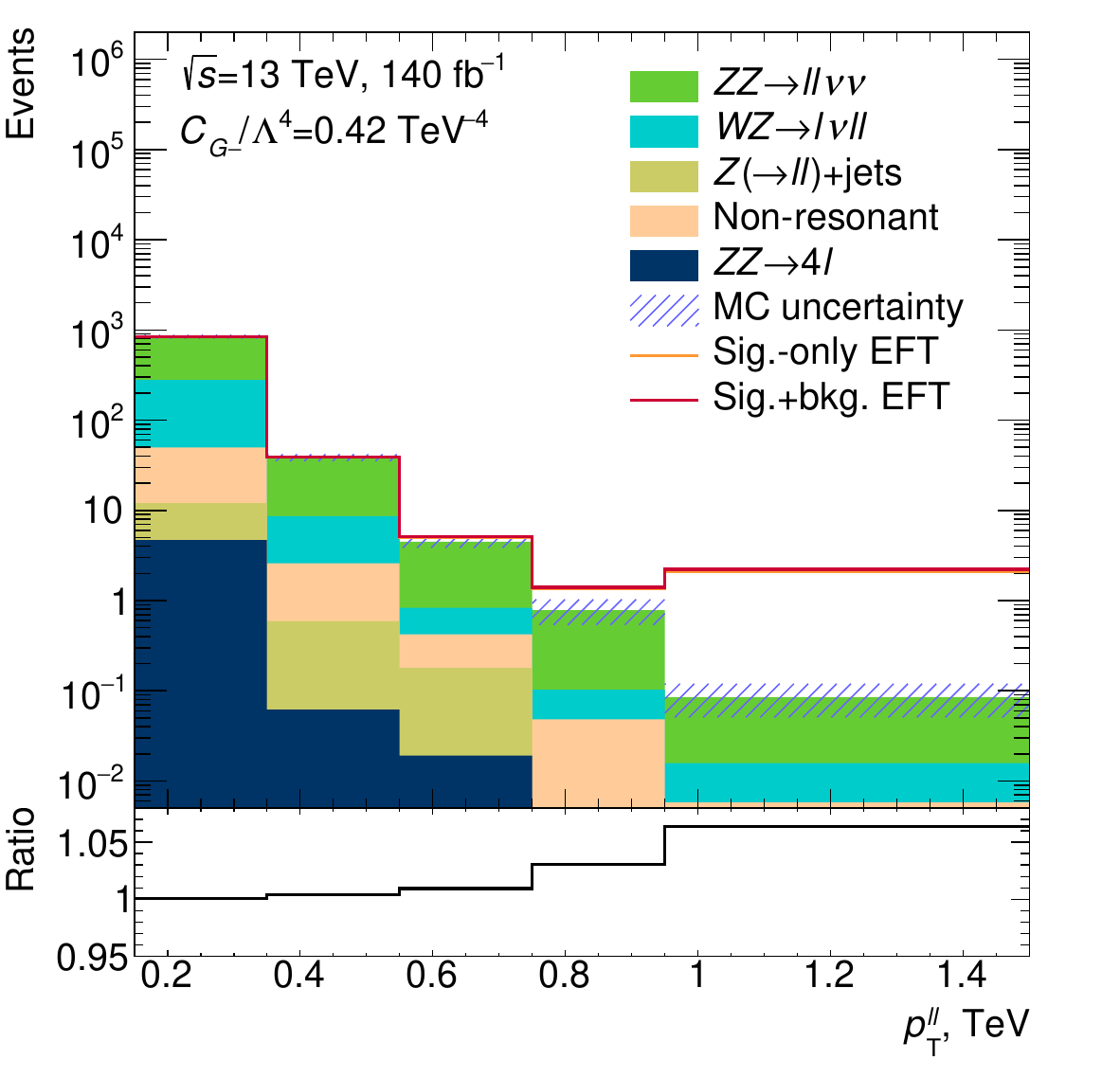}
\includegraphics[width=0.3\textwidth]{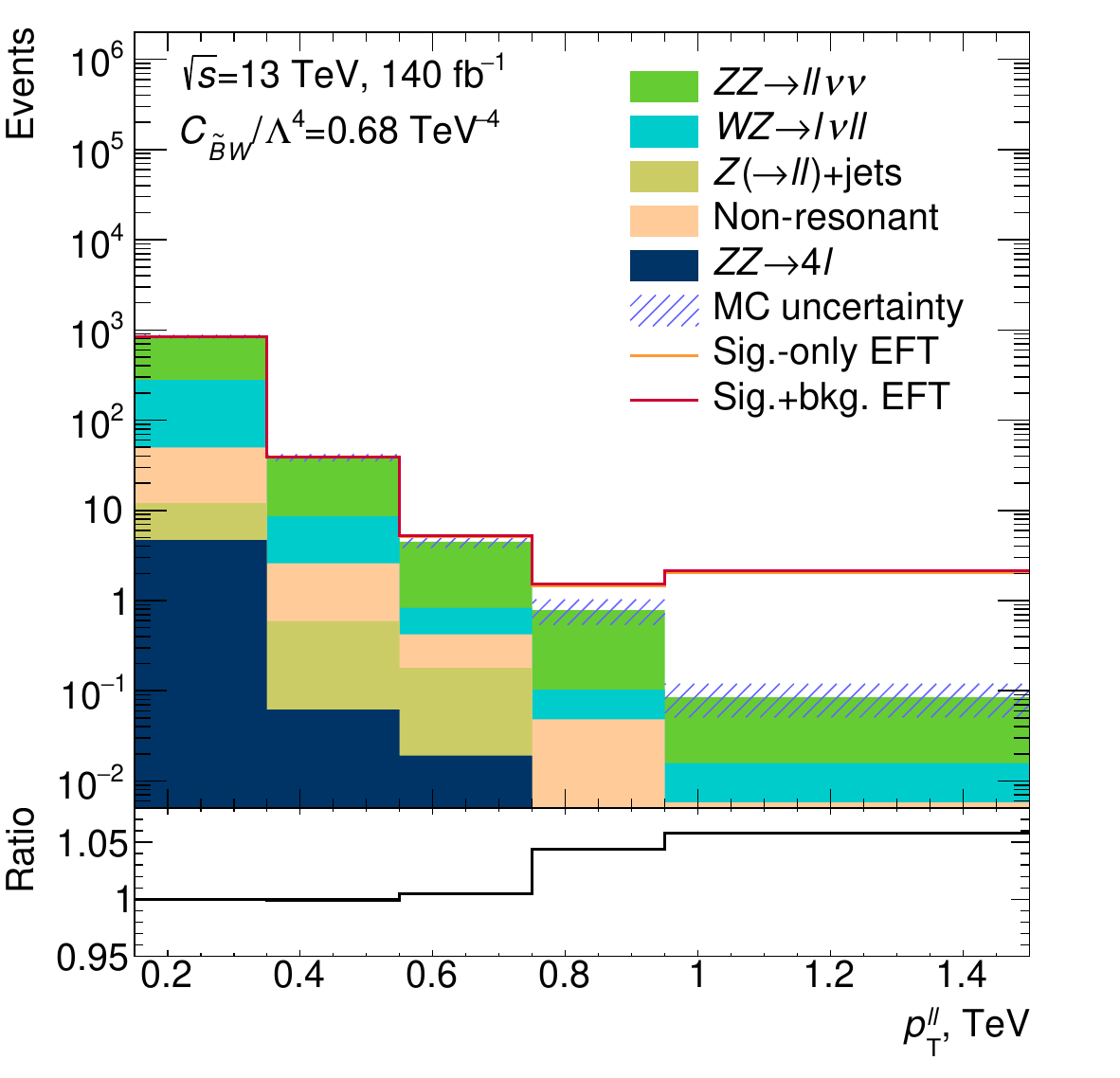}
\caption{$p_\text{T}^{\ell\ell}$ distributions of expected event yields. Filled histogram shows contributions from SM signal and each background processes. Orange (red) line represents expected event yields if one Wilson coefficient is non-zero for the case if BSM contributions from signal process only (signal and all background processes) are taken into account: $\cgp$~(left), $\cgm$~(center) and $\cbtw$~(right). At the lower panel ratio of red and orange lines is presented.}
\label{fig:ptll_sm}
\end{figure}

\begin{figure}[h!]
\centering
\includegraphics[width=0.3\textwidth]{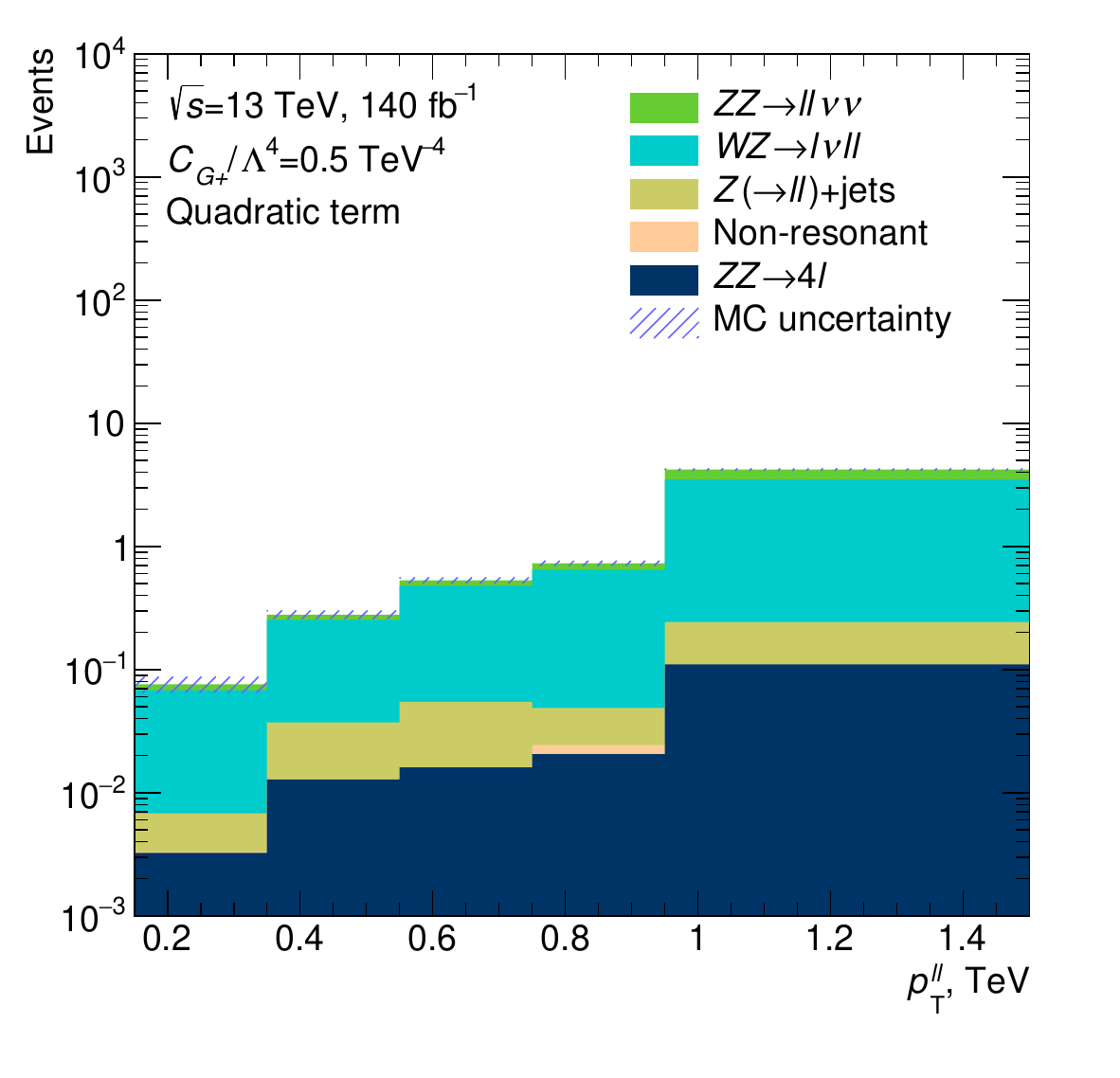}
\includegraphics[width=0.3\textwidth]{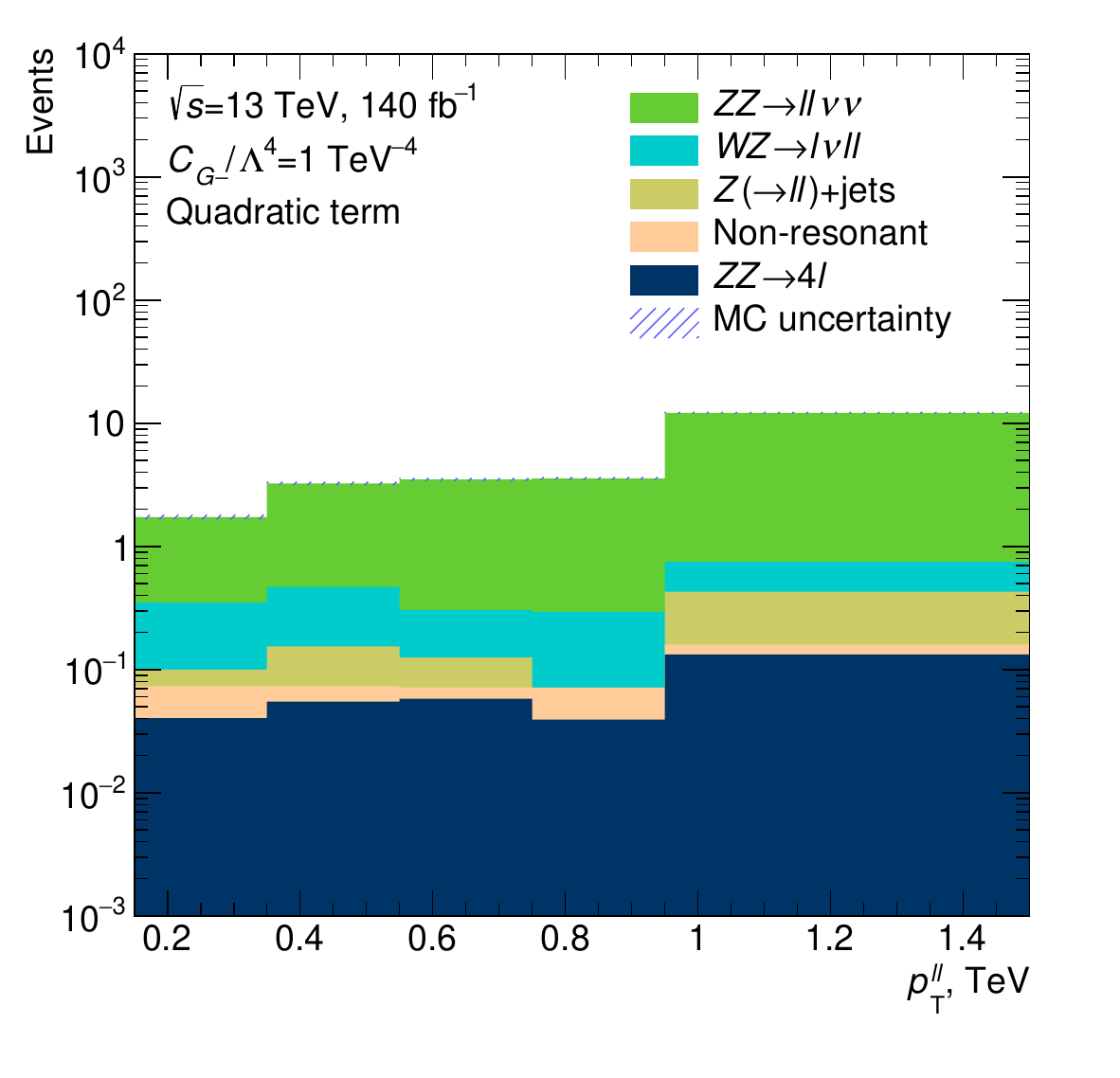}
\includegraphics[width=0.3\textwidth]{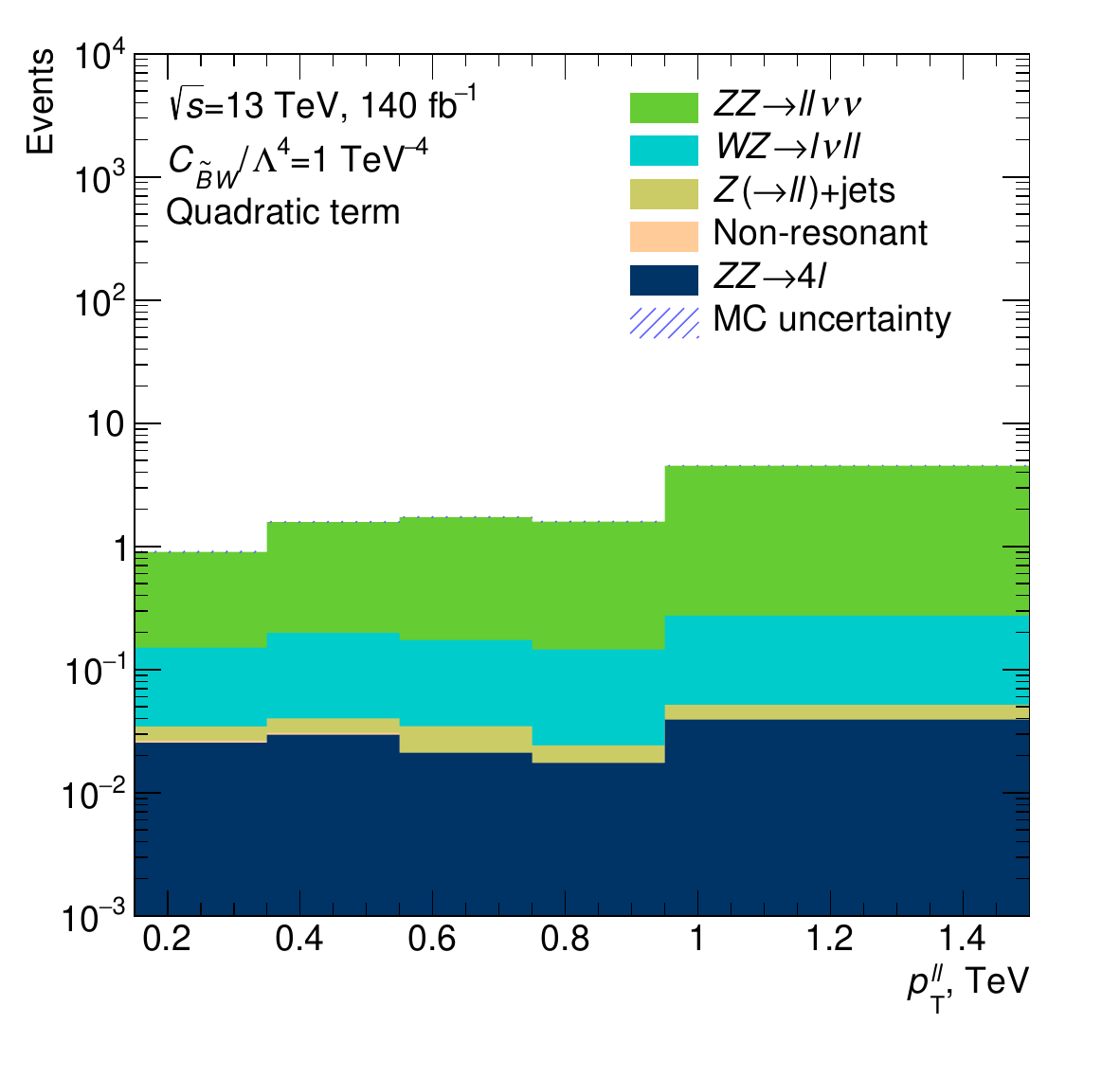}
\caption{$p_\text{T}^{\ell\ell}$ distributions of expected event yields for quadratic BSM term only, showing contribution from signal and each background processes, for coefficients $\cgp$~(left), $\cgm$~(center) and $\cbtw$~(right).}
\label{fig:ptll_quad}
\end{figure}

\begin{figure}[h!]
\centering
\includegraphics[width=0.3\textwidth]{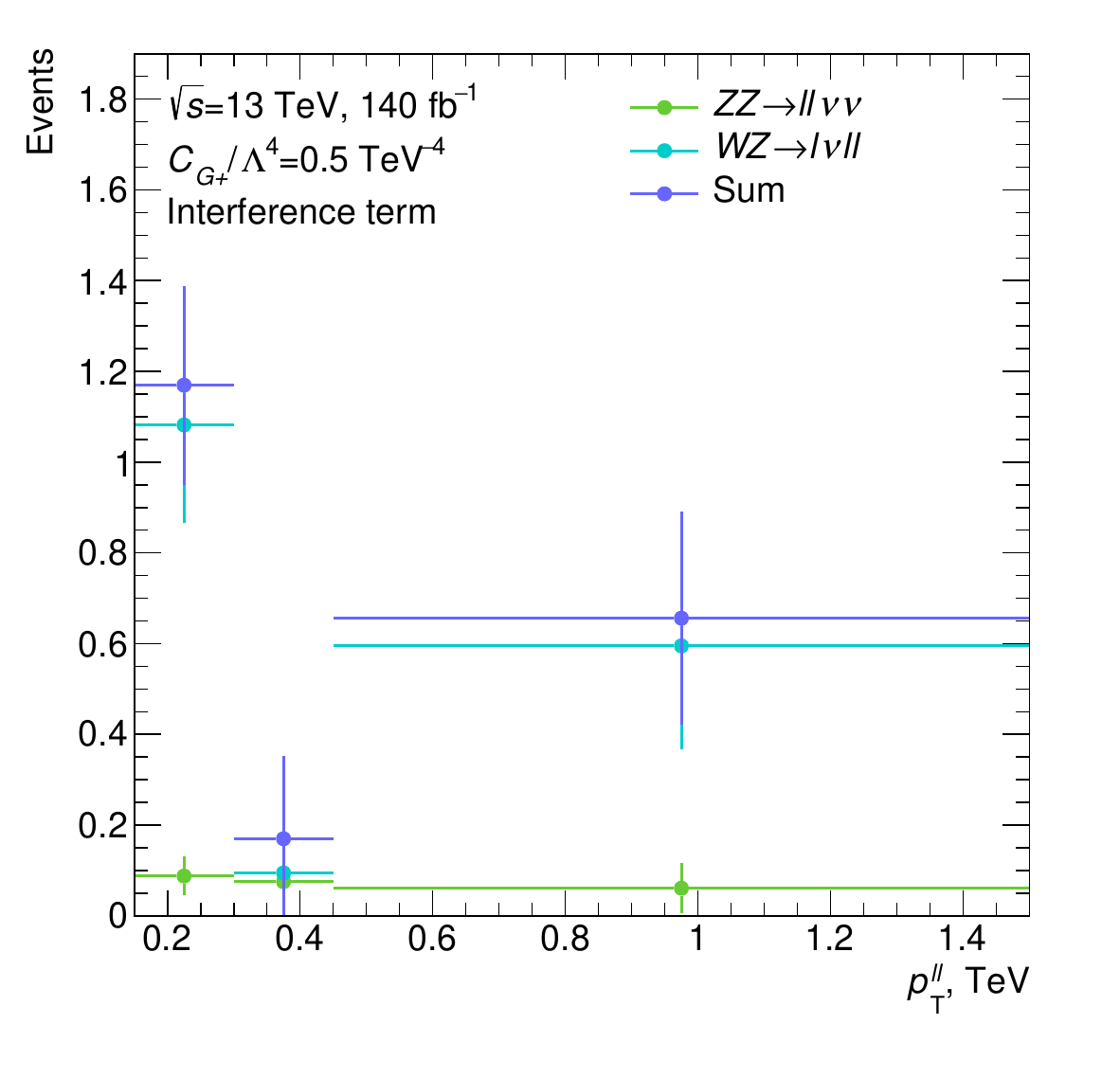}
\includegraphics[width=0.3\textwidth]{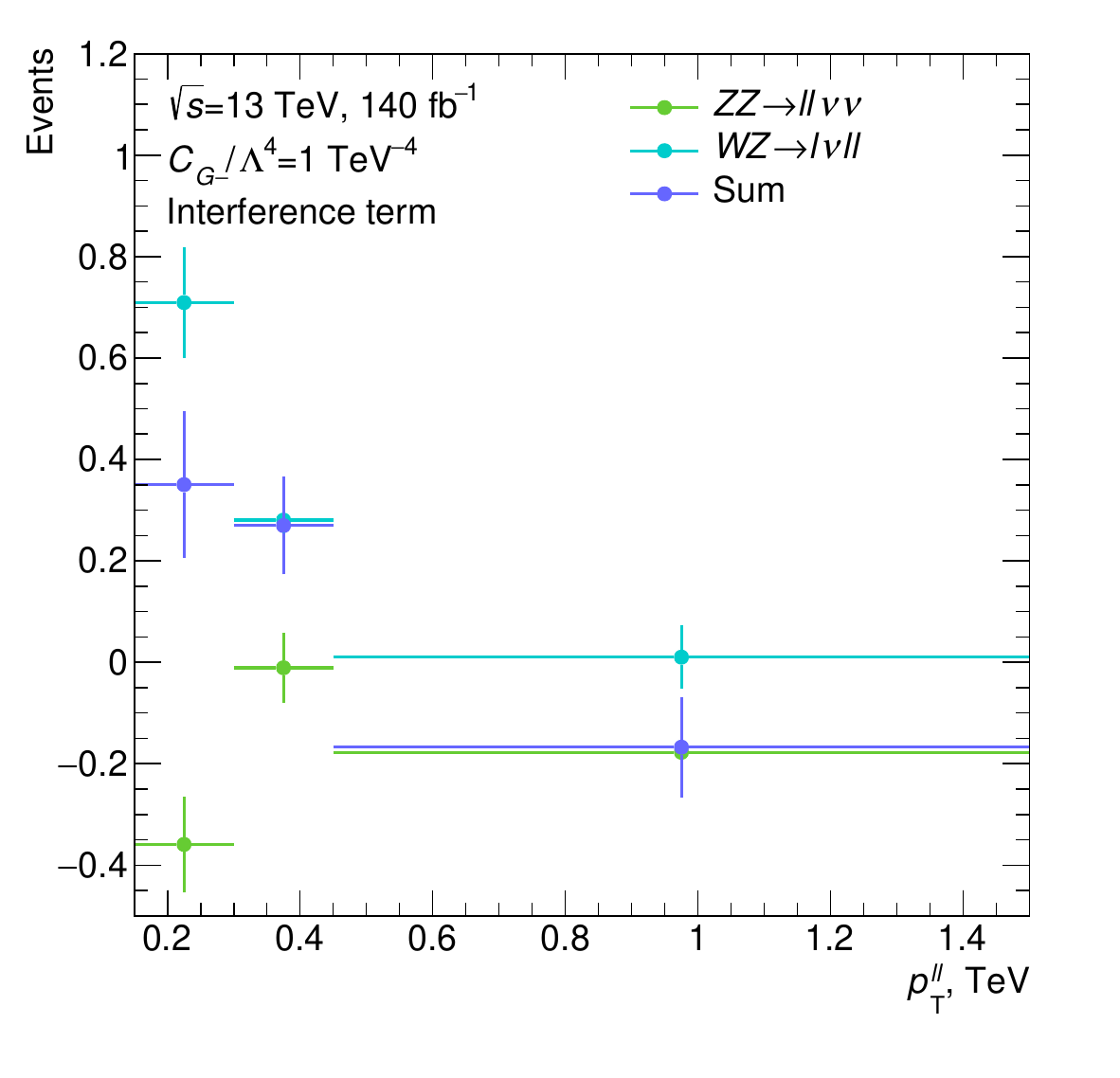}
\includegraphics[width=0.3\textwidth]{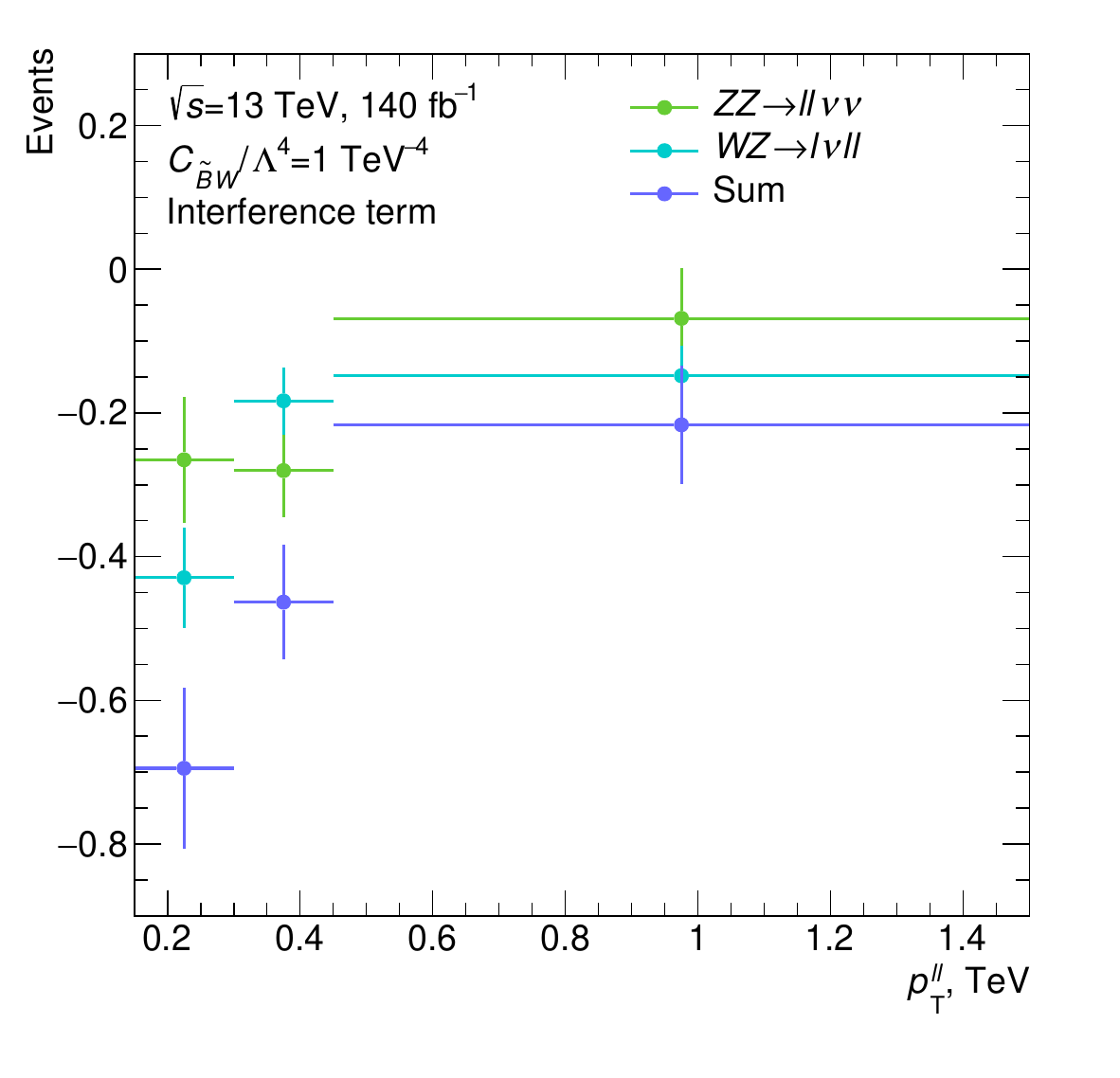}
\caption{$p_\text{T}^{\ell\ell}$ distributions of expected event yields for interference BSM term only, showing contribution from signal and main background process, for coefficients $\cgp$~(left), $\cgm$~(center) and $\cbtw$~(right).}
\label{fig:ptll_int}
\end{figure}

For setting limits on the Wilson coefficients, momentum of two leptons $p_\text{T}^{\ell\ell}$ is used as a sensitive variable due to the large sensitivity of this variable to anomalous couplings, especially in high-$p_\text{T}^{\ell\ell}$ region~\cite{ATLAS:2019xhj, CMS:2015qgb}. Distributions by this variable for three different Wilson coefficients with illustrations of contributions from each SM background are presented in Fig.~\ref{fig:ptll_sm}. Fig.~\ref{fig:ptll_quad} shows contributions to the quadratic term from signal and each background processes. Fig.~\ref{fig:ptll_int} shows contributions to the interference term from signal and main ($\wzlvll$) background processes. Due to the large Monte Carlo uncertainties, binning of Fig.~\ref{fig:ptll_int} was changed and backgrounds except $\wzlvll$ were dropped. Conventional method for setting the limits on the Wilson coefficients uses BSM contributions from the signal process only, whereas in general case EFT signal is composed of contributions from different processes.

Usually main backgrounds are estimated from data, using special control regions, enriched by the corresponding events. Therefore, any estimation of the background (its signal strength) already includes some BSM contributions, and method of composite anomalous signal can become not valid. However, the method works correctly if the control region is enriched by the SM background events and has small sensitivity to the anomalous couplings. This can be reached if the control region is splitted by the variable that is not sensitive to anomalous couplings. Moreover, it is possible to use a sensitive variable but in a non-sensitive region, e.g. $p_\text{T}^{\ell\ell}$ in low-$p_\text{T}^{\ell\ell}$ region, that is enriched by the SM events and is not enriched by the anomalous events, as can be seen in Fig.~\ref{fig:ptll_sm}. Finally, the method is obviously valid if the background is estimated directly from the Monte Carlo simulation. Often these conditions are satisfied in real analyses~\cite{CMS:2015qgb, ATLAS:2019xhj, ATLAS:2022nru}, so the composite anomalous signal can be used.

\section{Statistical limit-setting procedure and results}
Limits on the Wilson coefficients are set using frequentist statistical method with likelihood-ratio-based test statistic~\cite{Cowan:2010js}, defined as
\begin{equation}
    t_{\mmu} = -2 \ln \frac{L(\mmu,\hhtth(\mmu))}{L(\hmmu,\htth)},
\end{equation}
where $\mmu$ and $\tth$ are vectors of parameters of interest and nuisance parameters respectively, $L(\mmu,\tth)$ is a likelihood function, $(\hmmu,\htth)$ is its global maximum and $\hhtth(\mmu)$ maximizes the likelihood function for a fixed $\mmu$. Confidence region for the confidence level (CL) of 95\% is defined as a region in $\mmu$-space, that satisfies the condition
\begin{equation}
    p_{\mmu} = \int\limits_{t_{\mmu}^\text{obs}}^\infty f(t_{\mmu} | \mmu) \, \text{d} t_{\mmu} > 0.05,
\end{equation}
where $t_{\mmu}^\text{obs}$ is the observed value of the test statistic and $f(t_{\mmu} | \mmu)$ is the distribution of the test statistic. The latter is assumed to be asymptotic, i.e. chi-squared with one or two degrees of freedom for one- or two-dimensional limits respectively, according to the Wilks' theorem~\cite{Wilks:1938dza}.

Likelihood function consists of a product of Poisson distributions for each bin and Gaussian constraints of the nuisance parameters. For linear~+~quadratic model binning as in Figs.~\ref{fig:ptll_sm} and~\ref{fig:ptll_quad} is used, whereas for linear model binning is reduced to the one from Fig.~\ref{fig:ptll_int}. Separate nuisance parameters are created for Monte Carlo uncertainties in each bin, and additional systematic uncertainty of 10\% is applied in order to make the limits more real, but do not contribute significantly to the results. Since this study does not use experimental data, SM prediction is used as data to obtain the expected limits.

The one-dimensional limits on the Wilson coefficients in linear~+~quadratic model are set for three cases: A) if only signal process BSM contributions are accounted, B) if BSM contributions from only one background are added to the signal one, C) if contributions from signal and all the backgrounds are accounted simultaneously. Table~\ref{tab:Results1DQuad} contains the limits for case A and impact from each background BSM contribution on the limits. This impact shows how much the confidence interval from the case B is stronger than the one from the case A. Finally, the limits for the case C and impact from all the background EFT contributions are presented. It can be seen that the main improvement comes from $\wzlvll$ background, whereas other backgrounds have negligible effect on the limits on the Wilson coefficients. The exceptions are $\ogm$ and $\obb$ operators, since $\wzlvll$ production has small and zero sensitivity to them respectively. The most significant improvement is obtained for coefficients $\cgp$ and $\cww$ due to the large sensitivity of $\wzlvll$ production to the new physics described by corresponding operators.
\begin{table}[h!]
    \centering
    \caption{Impact of the BSM background contributions on the one-dimensional limits on six Wilson coefficients in linear~+~quadratic model.}
    \label{tab:Results1DQuad}
    \begin{scriptsize}
    \begin{tabular}{|c|c|c|c|c|c|c|c|}
        \hline
        \multirow{2}{*}{Coef.} & Limits [TeV$^{-4}$], & \multicolumn{4}{c|}{Impact from separate bkgs.} & Limits [TeV$^{-4}$], & \multirow{2}{*}{Total bkg. impact} \\ \cline{3-6}
        & sig.-only EFT & $\wzlvll$ & Non-res. & $\zllj$ & $\zzllll$ & all bkgs. EFT & \\ \hline
        $\cgp$ & [-0.122; 0.122] & 57.5\% & 0 & 8.2\% & 0.7\% & [-0.051; 0.050] & 58.8\% \\
        $\cgm$ & [-0.422; 0.423] & 1.6\% & 0.2\% & 1.0\% & 0.6\% & [-0.408; 0.410] & 3.3\% \\
        $\cbtw$ & [-0.679; 0.686] & 2.8\% & 0 & 0.2\% & 0.5\% & [-0.654; 0.665] & 3.4\% \\
        $\cbw$ & [-1.57; 1.55] & 3.4\% & 0 & 0.3\% & 0.5\% & [-1.50; 1.49] & 4.1\% \\
        $\cbb$ & [-0.836; 0.839] & 0 & 0 & 0.2\% & 0.6\% & [-0.829; 0.833] & 0.8\% \\
        $\cww$ & [-1.31; 1.28] & 11.1\% & 0 & 0.5\% & 0.5\% & [-1.15; 1.13] & 11.7\% \\ \hline
    \end{tabular}
    \end{scriptsize}
\end{table}

Table~\ref{tab:Results1DLin} presents the one-dimensional limits on the Wilson coefficients of $CP$-conserving operators in linear model. They were set without accounting for background BSM contributions and with accounting ones from $\wzlvll$ background. Interference terms of other backgrounds are very small and, therefore, are not taken into account. The limits in linear model are much weaker than the ones in linear~+~quadratic model, since at the current experimental sensitivity quadratic term dominates. However, the limits can be significantly improved by composite anomalous signal, background impact on the limits is also presented in Table~\ref{tab:Results1DLin}. Note that in general case limits in linear model can be improved as well as worsened. For example, interference terms of $\zzllvv$ and $\wzlvll$ productions for $\ogm$ operator have different signs, it can be seen in Fig.~\ref{fig:ptll_int}. Therefore, usage of different binning and selection compared to the ones in this work can lead to the worsening of the limits.
\begin{table}[h!]
    \centering
    \caption{Impact of the BSM background contributions on the one-dimensional limits on six Wilson coefficients in linear model.}
    \label{tab:Results1DLin}
    \begin{tabular}{|c|c|c|c|}
        \hline
        {Coef.} & Limits [TeV$^{-4}$], sig.-only EFT & Limits [TeV$^{-4}$], sig.+bkg. EFT & {Bkg. impact} \\ \hline
        $\cgp$ & [-147; 229] & [-9; 11] & 94.4\% \\
        $\cgm$ & [-110; 94] & [-57; 60] & 42.9\% \\
        $\cbtw$ & [-77.5; 66.2] & [-37; 32] & 56.2\% \\ \hline
    \end{tabular}
\end{table}

\begin{figure}[h!]
\centering
\includegraphics[width=0.3\textwidth]{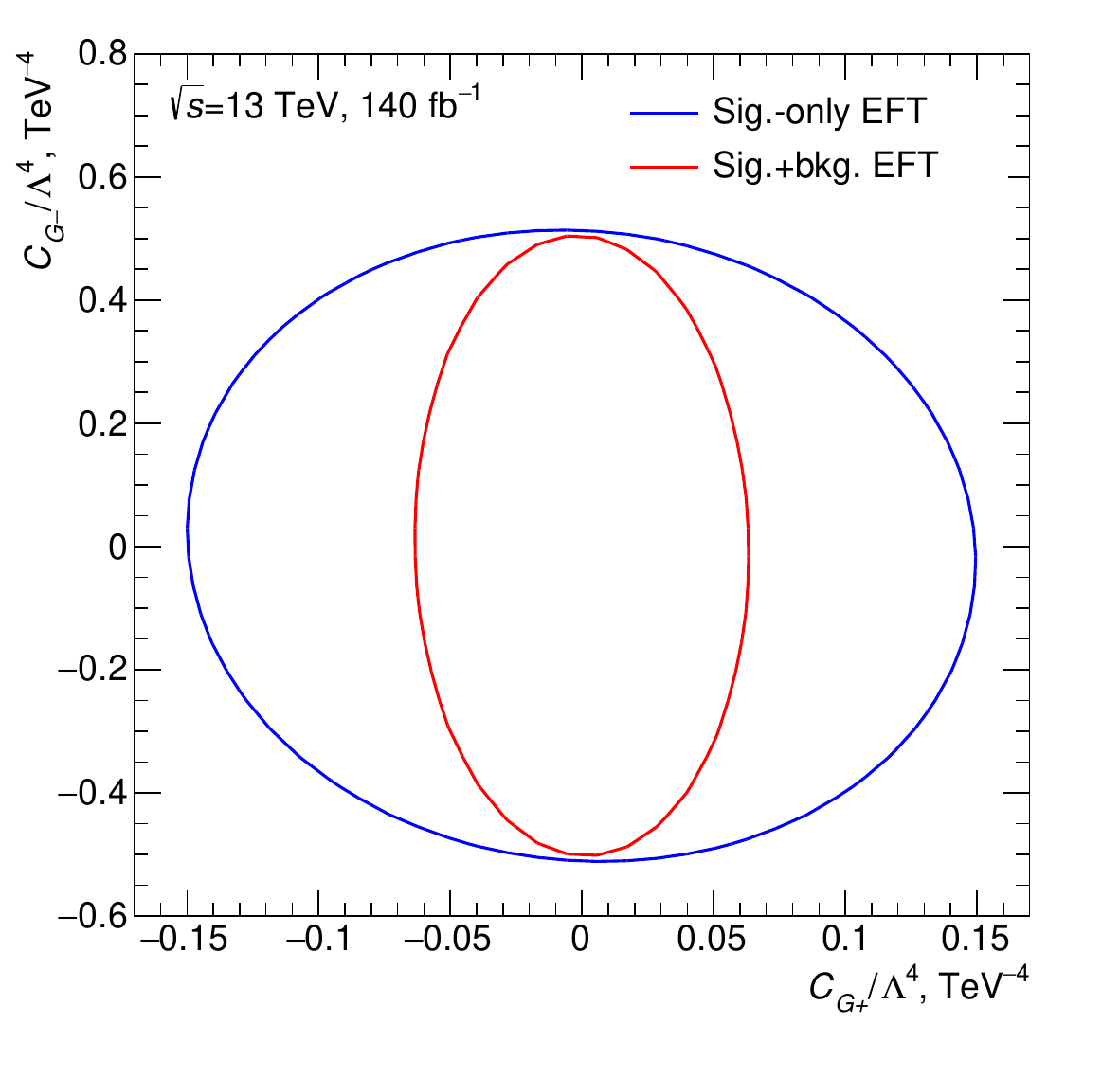}
\includegraphics[width=0.3\textwidth]{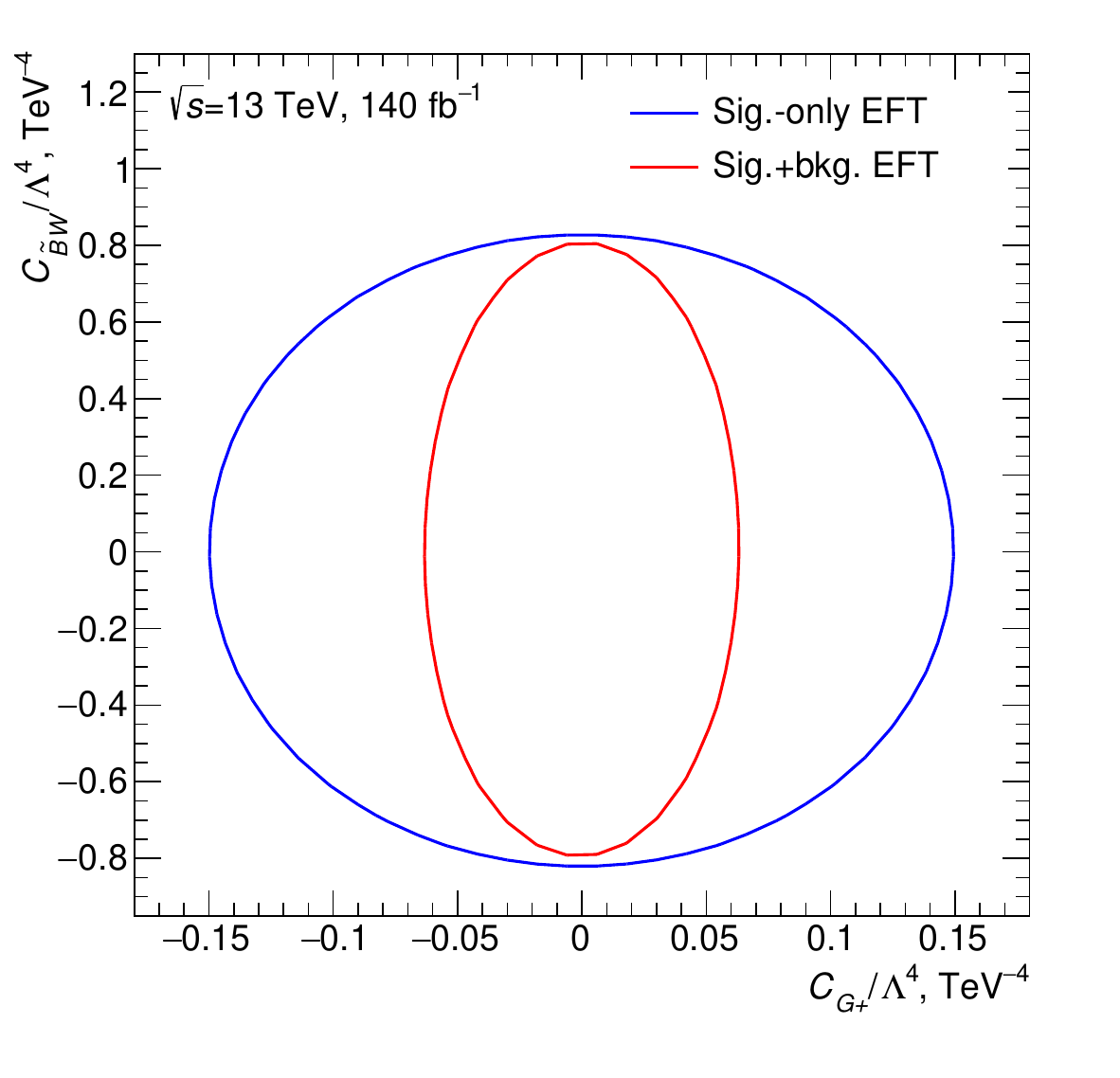}
\includegraphics[width=0.3\textwidth]{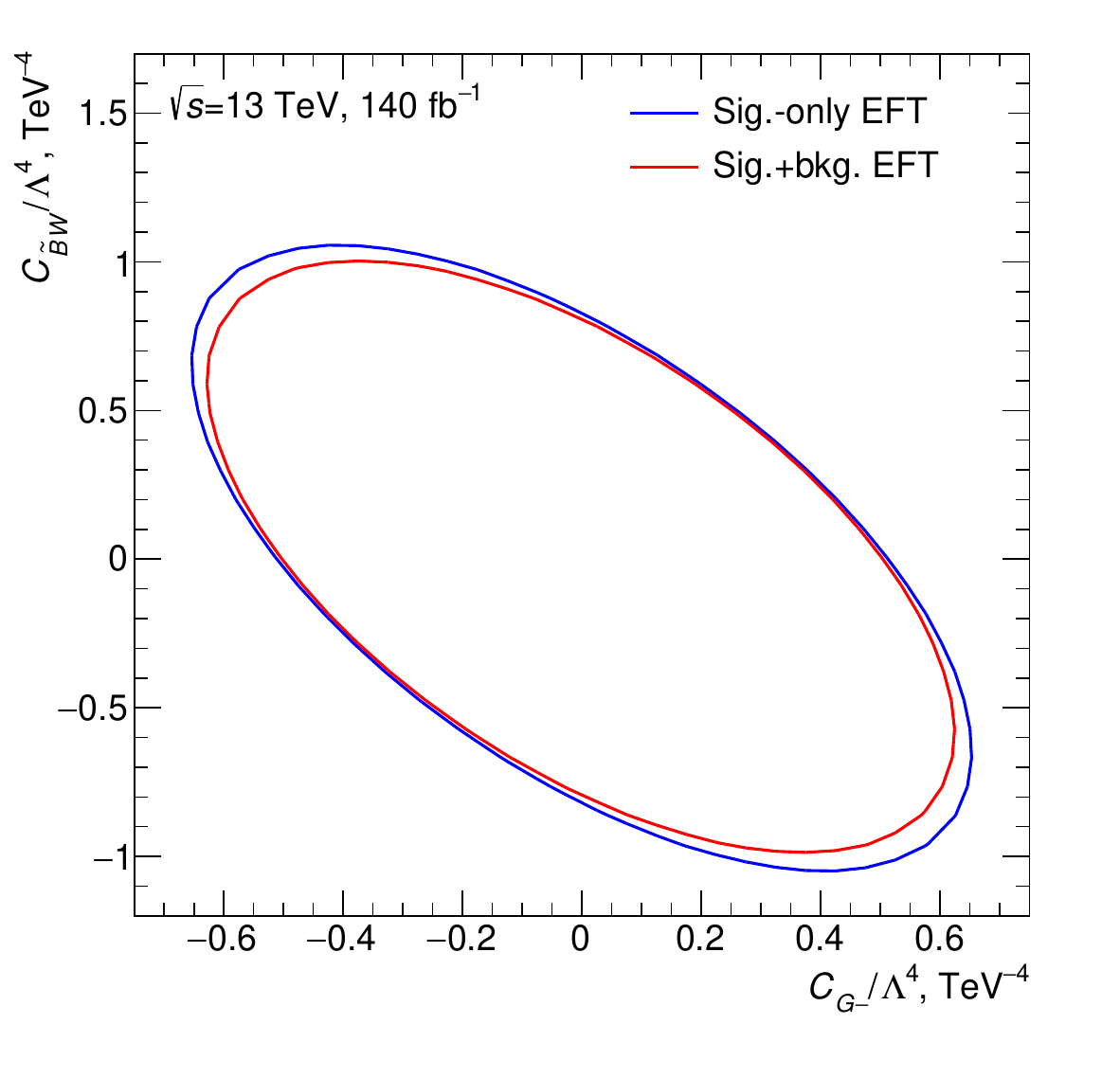}

\includegraphics[width=0.3\textwidth]{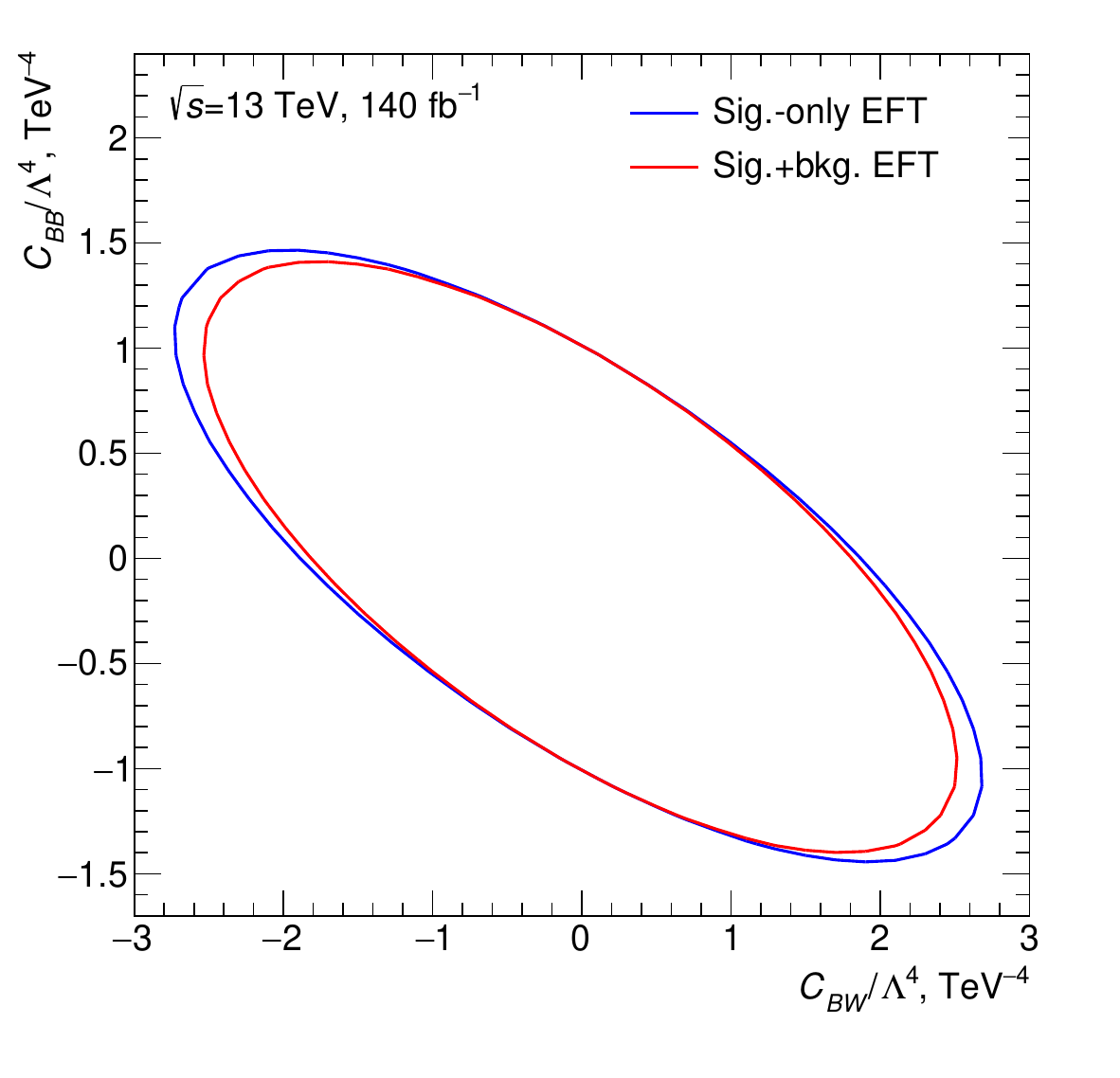}
\includegraphics[width=0.3\textwidth]{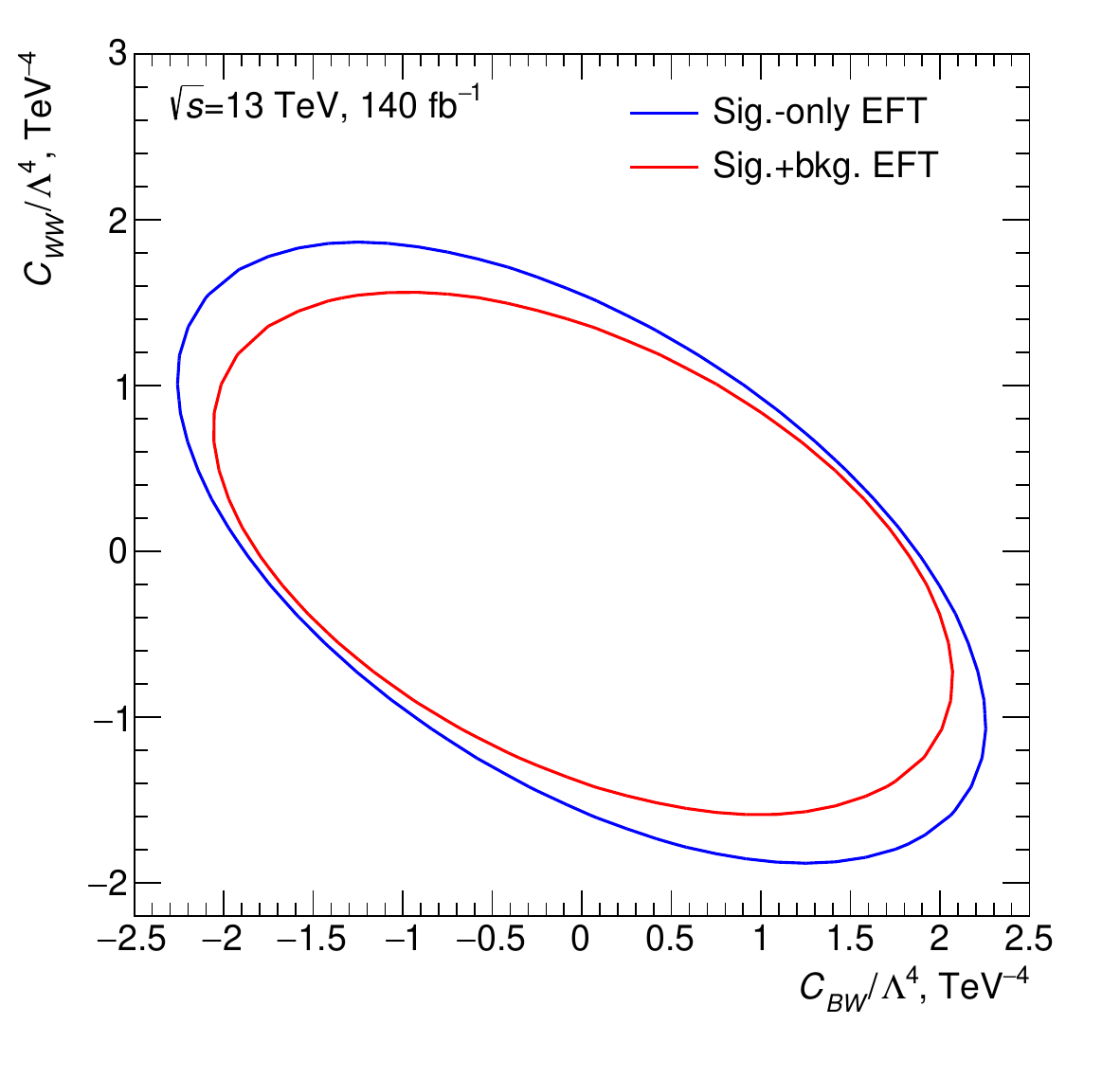}
\includegraphics[width=0.3\textwidth]{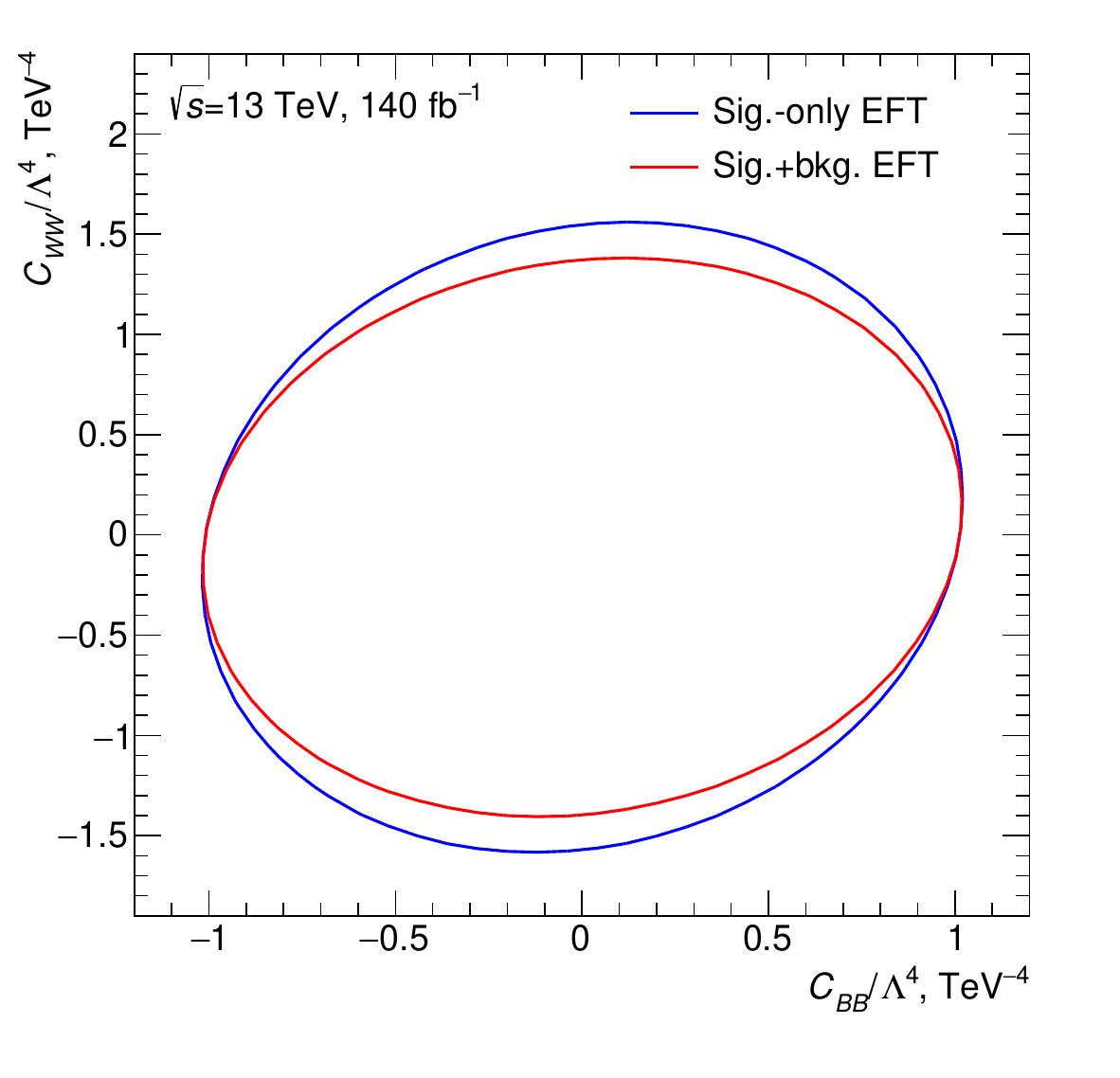}
\caption{Two-dimensional 95\% CL limits on six pairs of the Wilson coefficients. Blue (red) contours were obtained under usage of the anomalous contributions from $\zzllvv$ ($\zzllvv+\wzlvll$) production.}
\label{fig:Results2D}
\end{figure}

Additionally, two-dimensional limits are also set with and without BSM contributions from $\wzlvll$ background, dropping BSM contributions from other backgrounds due to their smallness. $CP$-even and $CP$-odd operators have zero interference (cross-term) in this study. Thus, two-dimensional limits are set on six pairs of the Wilson coefficients, and they are presented in Fig.~\ref{fig:Results2D}. As it was mentioned above, for some coefficient pairs cross-term leads to the change of the sensitivity compared to the one-dimensional limits. Area outside the presented contours is excluded at 95\% CL. The improvement of the two-dimensional limits due to taking into account $\wzlvll$ BSM contributions varies from 6.7\% for $\cbw$ vs $\cbb$ contour to 59.1\% for $\cgp$ vs $\cbtw$ contour.

\section{Conclusion}
This paper presents a study of BSM contributions of background processes in the search of anomalous neutral triple gauge couplings using $\zzllvv$ production in $pp$ collisions at $\sqrt{s}=13$~TeV with a dataset of 140~fb$^{-1}$. Six dimension-eight operators describing corresponding couplings were considered. The results obtained in linear~+~quadratic model showed that, generally, only $\wzlvll$ background changes the limits on the Wilson coefficients significantly, and the most significant improvement was found for $\cgp$~(57.5\%) and $\cww$~(11.1\%) coefficients. Additionally, the improvement of the limits on the Wilson coefficients in linear model was derived and is of 42.9\%-94.4\% depending on the operator. Finally, two-dimensional limits in linear~+~quadratic model were considered, and the confidence regions became tighter up to 59.1\%.

Wilson coefficients can be turned into the model-dependent parameters. Therefore, experimental limits on these coefficients allow constraining some parameters of BSM theories. So, different possibilities of improvement of the limits on the Wilson coefficients, like the presented in this work method of composite anomalous signal, should be studied and used in the analyses of the experimental data.

\section*{Acknowledgements}
This work was supported by the Russian Science Foundation under grant 21-72-10113.

\bibliographystyle{JHEP}
\bibliography{references}

\end{document}